\documentclass[preprint]{aa} 
\usepackage{graphicx}

\newcommand{\be}{\begin{equation}} 
\newcommand{\ee}{\end{equation}}
\newcommand{\etal}{{\it et al.\ } } 

\newcommand{\gne}[2]{\\vphantom{S}^{\raise-0.5pt\hbox{$\scriptstyle#1$}}_{\raise0.5pt\hbox{$\scriptstyle#2$}}}

\newcommand{\unit}[1]{\,{\rm #1}}
 
\newcommand{\ghz}{\unit{GHz}}

\newcommand{\mjy}{\unit{mJy}}

\newcommand{\ergsh}{\unit{erg\,sec^{-1}\,Hz^{-1}}}
\newcommand{\aop}{\alpha_{\rm op}} 
\newcommand{\ar}{\alpha_{\rm r}}

\newcommand{\hubble}[1]{H_0={#1}\unit{km\,sec^{-1}\,Mpc^{-1}}}

\newcommand{\labfig}[1]{\label{fig:#1}}
\newcommand{\labtab}[1]{\label{tab:#1}}
\newcommand{\labsecn}[1]{\label{sec:#1}}

\newcommand{\tab}[1]{Table~\ref{tab:#1}}
\newcommand{\fig}[1]{Figure~\ref{fig:#1}}

\titlerunning{Radio Emission from AGN}

\begin{document} 
\title{Radio Emission from AGN detected by the VLA FIRST Survey}
\author{Yogesh Wadadekar\inst{1,2}}
\institute{Institut d'Astrophysique de Paris, 98bis Boulevard Arago, 75014 
Paris, France \and Space Telescope Science Institute, 3700 San Martin Drive, 
Baltimore, MD 21218, USA}
\date{}

\abstract{
Using the most recent (April 2003) version of the VLA FIRST survey
radio catalog, we have searched for radio emission from $>2800$ AGN taken
from the most recent (2001) version of the Veron-Cetty and Veron AGN
catalog. These AGN lie in the $\sim$9033 square degrees of sky
already covered by the VLA FIRST survey.  Our work has resulted in
positive detection of radio emission from 775 AGN of which 214
are new detections at radio wavelengths.
\keywords{galaxies: active -- methods: statistical -- catalogs --
surveys} }
\maketitle
\section{Introduction}

Systematic radio observations of samples of active galaxies (AGN)
 have a long history (throughout this paper we use the term AGN to designate all active
galactic nuclei, except quasars). De Bruyn \& Wilson (1976, 1978)
surveyed a sample of 43 Markarian Seyferts and found that Seyfert 2
galaxies have on the average a higher radio luminosity than Seyfert 1
galaxies. Ulvestad \& Wilson (1984) used high
resolution data from the VLA on a radio flux limited sample of 79
Markarian Seyferts and a distance limited sample of 25 Seyfert
galaxies. They also found that Seyfert 2 galaxies are systematically
more luminous than Seyfert 1 galaxies and that a weak correlation
exists between sizes and powers of nuclear radio sources in
Seyferts. On average, Seyfert 2 radio sources were found to be larger than those of
Type 1 Seyferts. However, in a subsequent study,
Ulvestad \& Wilson (1989) found that their sample of Seyfert 2's could
have been biased in radio luminosity.  Their subsequent analysis of a
distance limited sample of 27 Seyfert galaxies that was free of this
bias showed that the earlier conclusions were not statistically
significant. This new result supported the finding of
 Edelson (1987) that there was no significant
difference between the radio luminosities of the Seyfert 2 and Seyfert
1 galaxies in his magnitude limited sample of 42 Seyferts. More
recently, Ulvestad \& Ho (2001) found that Seyfert 1 galaxies host
somewhat stronger radio sources than Seyfert 2 galaxies at 6 and 20
cm, particularly among the galaxies with the weakest nuclear
activity. They used a sample of 45 Seyfert galaxies from the Palomar
spectroscopic survey of nearby galaxies.
 
The aim of many of the above studies was to use statistical properties
of radio sources in these galaxies to draw conclusions about their
physical character and to test the unification scheme. According to
the unification scheme for AGN, at a given intrinsic luminosity,
\emph{all} other properties of AGN from spectroscopic classification
to VLBI component speeds can be ascribed to orientation effects.
 
A straightforward test of the unification hypothesis is to see if a
property that is \emph{not} orientation dependent has similar values
and distribution for different classes of AGN. Assuming that the
standard unification scheme for AGN is valid, radio and FIR
luminosities of AGN are expected to be orientation independent, except
over a small angle where Doppler boosting of radio jets occurs. This
is because neutral gas and dust surrounding the radio source itself
cannot obscure it, and even ionized gas (found in nuclear starbursts)
is transparent at radio frequencies \( \nu >3 \) GHz,  for AGN with low to moderate emission measure. In fact, the
most compact nuclear starbursts are transparent only at GHz radio and
far infrared (FIR) wavelengths. At other wavelengths, absorption of
the radiation by the intervening matter surrounding the nucleus is
significant and it is very difficult to measure the intrinsic
luminosity.  In principle, the unification scheme can be tested by
radio observations of different kinds of AGN.

In the standard unification scheme for Seyferts, it is expected that
Seyfert 1 and Seyfert 2 galaxies should have similar total radio power
(since the radio emission is largely unobscured at centimeter
wavelengths). Also, if the predominant radio emission is from jets,
radio jets from Seyfert 1 would be foreshortened as they are viewed
closer to end-on. Statistically, this implies that Seyfert 1 radio
source sizes should be smaller than Seyfert 2 sizes. Schmitt et
al. (2001) have found this effect in a sample of Seyfert galaxies
selected at 60 $\mu$m.

All the previous studies of radio emission in AGN have suffered due to
their small sample sizes (typically 50 galaxies). Some of them have
also been affected by subtle selection biases that
were only discovered many years later. In this work, we approach the
problem from a different perspective. We use data from the recently
completed VLA FIRST survey to look for radio emission from all known
AGN, lying in the VLA-FIRST survey area. Our sample size is an order
of magnitude larger than typical sample sizes in previous
studies. However, it is by no means a well defined complete sample,
but is motley collection of active galaxies. It is a supersample of
all previous samples, subject to all the selections biases and
incompleteness of each of these. It is therefore, impossible to use
our sample for construction of radio luminosity functions. The large
sample size, however, may make it possible to recognize significant
statistical trends in the data.
 
Our AGN sample is drawn from the most recent (10th) edition of the
Veron-Cetty and Veron (2001) catalog, which is a complete survey of
the literature and lists all quasars and active nuclei known to the
authors of the catalog prior to May 1, 2001. They define a quasar as a
starlike object, or object with a starlike nucleus, brighter than
absolute magnitude \( M_{B}=-23 \) (using \( \hubble {50} \), \(
q_{0}=0 \)). Objects fainter than this limit appear in Table 3 of the catalog,
which we have used in this work (hereafter VV01). Some of the objects
listed in this table would move to Table 1 (which lists quasars) and
vice versa for a different choice of cosmological
parameters. Variability and the size of the diaphragm used for the
apparent magnitude measurement may have a similar effect, as the
contribution of the underlying host galaxy may be significant for the
faint quasars. Such a criterion has the additional shortcoming that at
higher redshifts the AGN luminosity function is artificially
truncated. However, these caveats in sample identification are offset
by the large size of the sample - 5751 active galaxies are listed in
Table 3 of the catalog, of which about 2840 lie in the area
covered by the FIRST survey.
 
Deep, high resolution, large area radio surveys provide the most
efficient means to detect radio emission from known AGN. The recently
completed VLA {\em Faint Images of the Radio Sky at Twenty
centimeters} (FIRST) survey (Becker \etal 1995; for more upto date
information see the FIRST survey homepage at {\em
http://sundog.stsci.edu/}) combines a large sky coverage with a low
flux limit of $1 \mjy$ at 20 cm. This survey now covers $>9000$ square
degrees mostly around the North Galactic Cap with a small equatorial
strip, the same area of the sky now being surveyed by the Sloan
Digital Sky Survey (SDSS; http://www.sdss.org/). To date, data for
approximately 9033 square degrees of sky have been released. The FIRST
survey is now substantially complete. A small amount of additional
data to fill holes within the surveyed area may be acquired; but an
increase in the area covered is unlikely.

The FIRST survey has better sensitivity and resolution than its sister
survey the NRAO VLA Sky Survey (NVSS; Condon 1998), although it covers
a smaller area.  More importantly, the FIRST survey, which is being
carried out with the VLA in its B-configuration, has a resolution of 5
arcsec combined with excellent astrometric accuracy of $\sim1\arcsec$
(90\% error circle) and a 5 sigma sensitivity of $\sim1\mjy$. This
compares favorably with the D-array NVSS, which has a beam size of 45
arcsec and a 5 sigma sensitivity of $\sim2.4\mjy$.  FIRST has a
smaller beam size, a better resolution and more accurate astrometry
than NVSS.  It is therefore well suited for studies of radio emission
from compact sources such as quasars and AGN. A study of radio
emission from quasars using FIRST survey data has already been reported
(Wadadekar \& Kembhavi 1999).

\section{Radio/Optical Comparisons}

We compare the positions of AGN listed in VV01 to the positions of
radio sources in the FIRST radio source catalog (April 11, 2003 version
publicly available at {\em http://sundog.stsci.edu/}), and calculate
the angular separation between each AGN and each FIRST source. About
3.6\% of sources in the FIRST catalog have been tagged as possible
sidelobes of bright sources. Of these, $<$10\% are real sources and
considerably less than 1\% of the unflagged sources in the catalog are
sidelobes (White \etal 1997). We have excluded all such flagged sidelobe
sources from our cross correlation. We are then left with a total of
781,667 unflagged sources in the northern and southern strips, covering a
total area of about 9033 square degrees. On an average, there are
86.53 unflagged FIRST sources per square degree of sky.

In order to find coincidences between VV01 and FIRST sources, we adopt
the straightforward methodology used by Wadadekar \& Kembhavi
(1999). We begin with a search circle of radius of 600 arcsec centered
on each VV01 AGN, and look for FIRST radio sources within this
circle.  This search radius is clearly too large, given the uncertainties in the radio and optical positions. We choose this large trial radius only to robustly estimate chance coincidences. When there is more than one FIRST source in the search circle,
we tentatively count all such sources as matches. In \fig{sephistagn}
we show a histogram of the angular separation between the VV01 AGN
and the FIRST sources found within the search circles.

\begin{figure}
\resizebox{\hsize}{!}{\includegraphics{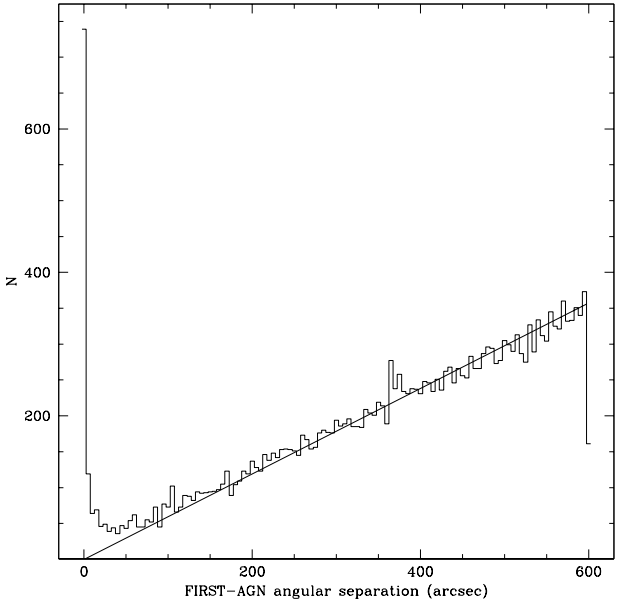}}
\caption{A histogram of the angular separation between the AGN and the corresponding FIRST source. The straight line is the number of chance matches expected, in annuli of radius shown on the X-axis having a width of 5 arcsec, if the FIRST sources were randomly distributed in the sky.}
\labfig{sephistagn}
\end{figure}

To see which of the radio sources found can be accepted as true
identifications, we estimate the AGN-FIRST source chance coincidence
rate for a random distribution of FIRST survey sources using a
technique identical to the one discussed in Wadadekar \& Kembhavi
(1999). The straight line in \fig{sephistagn} is the expected number
of chance coincidences between AGN and FIRST sources, in annuli of
radius shown on the abscissa and width 5 arcsec around the AGN. It
is seen from the figure that the expected number of chance
coincidences closely matches the number of actual coincidences for
radius $\geq$ 60 arcsec, indicating that most FIRST sources found more
than 60 arcsec away from the AGN are chance coincidences. On the other
hand, matches within 3 arcsec are almost all real (less than 0.05\%
chance identification probability).

The high resolution and excellent astrometric accuracy of the FIRST
survey makes it well suited for studies of compact sources.
Nevertheless, for our choice of cosmological parameters $(\hubble {50}
$ and $ q_{0 }=0.5)$, for an overwhelming majority of AGN in our
sample ($z<0.3$), an angular size of 3 arcsec corresponds to a spatial
scale of $<17$ kpc, comparable to the typical radius of a spiral
galaxy.  The astrometric error for FIRST sources depends on their
flux. Point sources at the detection limit of the survey have
positions accurate to better than 1 arcsec at 90\% confidence; 2 \mjy
  point sources in typically noisy regions have positions good to 0.5
arcsec. VV01 positions are compiled from the wide astronomical
literature on AGN and mean positional errors are impossible to
estimate. However, most of the brighter galaxies listed have positions
accurate to within an arcsec.  Given the combined root mean square
error in FIRST and VV01 positions, it is possible to claim that the
radio emission discussed here emanates from the galaxy concerned but
does not allow us to localise the emission within the
galaxy. Consequently, it is not possible to distinguish nuclear
emission from emission emanating from star-forming regions situated
away from the center of the galaxy. We use a search circle of radius 3
arcsec and count all matches found within this radius as \emph{true}
matches. All subsequent discussion about the radio properties of the
AGN only uses matches obtained with this search radius.

We have 775 AGN with a FIRST source located less than 3 arcsec away
from the centre of the AGN. Only 1 AGN (NGC 1068) has two FIRST
sources within three arcsec of its centre. Note that these ``AGN'' include all objects listed in Table 3 of VV01. Some of our detections are actually H-II regions, galaxies with unclassified spectra and LINERS (some of which may be starburst powered). There are 2068 non
detections, amongst the VV01 AGN covered by FIRST, and we assign an
upper limit of 1 mJy to their radio flux at 1.4 GHz. \tab{summagn}
provides a summary of our radio detections. \tab{summagn2} lists the
number of detections for the various types of AGN listed in
VV01.

The radio and optical properties of the AGN with detections are
summarized in Table 3.  The optical properties of the non-detections (upper limits) are
summarized in Table 4.  The columns are identical to those
in Table 3 except that the last three columns (which contain
radio data) are missing.\footnote{Tables 3 and 4 are only available in electronic form
at the CDS via anonymous ftp to \tt{cdsarc.u-strasbg.fr (130.79.128.5)}
or via \tt{http://cdsweb.u-strasbg.fr/cgi-bin/qcat?J/A+A/416/35}}

\begin{figure}
\resizebox{\hsize}{!}{\includegraphics{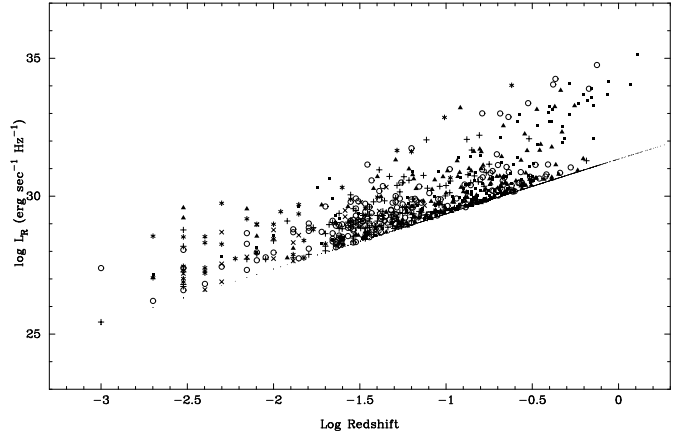}}
\caption{Radio luminosity as a function of redshift. The various symbols used correspond to the various kinds of AGN, as explained in the text. The lower locus of dots indicates the 1 mJy upper limits.}
\labfig{lrzagn}
\end{figure}

\section{Radio-optical luminosity corellations}

We have shown in \fig{lrzagn} a plot of the $5\ghz$ radio luminosity
against the logarithm of the redshift for the AGN. Throughout this paper, we compute luminosities and absolute magnitudes using optical and radio spectral indices, $\aop=\ar = -0.5$ with $L(\nu) \propto \nu^{+\alpha}$.  The non-detections
are shown as dots, and form the almost continuous lower envelope which
indicates the radio luminosity corresponding to a radio flux of
$1\mjy$ over the redshift range shown. The triangles indicate Seyfert
1 (S1,S1h,S1i,S1n), plus signs indicate intermediate type Seyfert galaxies
(S1.0,S1.2,S1.5,S1.8,S1.9; see the introductory
material accompanying VV01 for the exact definition of each AGN
category), the open circles indicate Seyfert 2 galaxies (S2), the
asterisks indicate Seyfert 3 (LINER) galaxies (S3,S3b,S3h), crosses
indicate H-II regions (H2 and HP), while the filled squares indicate
those sources for which the category is either not listed or deemed
suspect (S,S?,?,blank) in VV01. The same symbols are used in all
subsequent figures of this paper.

Seyfert 1 and Seyfert 2 galaxies span the entire redshift range shown
in the figure. However, H-II regions and LINERS are predominantly seen
at low redshifts. This is because the low average radio luminosity of
these objects makes them detectable above the FIRST flux limit only at
low redshifts. Another trend seen is that many of the highest redshift
AGN do not have a classification in VV01 (these are indicated by
filled squares). Many of these objects are in reality moderate
redshift quasars, which have been included in Table 3 of VV01, only
because they have absolute magnitude $M_{B}>-23$.

In \fig{lrabmagn} is shown a plot of the $5\ghz$ radio luminosity
against the absolute magnitude for the radio detections. For our
set of 775 AGN with radio detections, 688 have a V magnitude listed in VV01.
(NGC 1266 has a incorrect magnitude of 0.0, 14 objects have no
magnitude listed and 72 objects some other magnitude listed). We only
plot the points with a V magnitude available in this figure. There is a
clear correlation between the logarithm of the radio luminosity and
absolute magnitude. The formal linear correlation coefficient for the
Seyfert 1 is $-0.46$ while that for the Seyfert 2 is $-0.56$. Both
correlations are significant at the $>99.9$\% confidence level.

However, it is seen from \fig{abmzagn} and \fig{lrzagn} that mean
radio and optical luminosity both increase with redshift, which is due
to the existence of a limiting radio flux and apparent magnitude in
the surveys in which quasars are discovered. It is possible that the
observed correlation between radio and optical luminosities is mainly
due to the separate dependence of each luminosity on the redshift $z$. It is important to see if the correlation remains significant when
such an effect of the redshift on the observed correlation is taken
into account. This is done by evaluating a \emph{partial linear
correlation coefficient} as discussed in Wadadekar \& Kembhavi (1999).

For the Seyfert 1, the partial linear correlation coefficient is -0.13,
which is significant at only at the 97.6\% confidence level. This is at
variance with the findings of Ho \& Peng (2001) who reported a
significant correlation between {\em nuclear} radio and optical
emission for their sample of Seyfert 1. This discrepancy can be
explained by the very large range in absolute magnitude spanned by the
nuclear component of galaxies in their sample. The total (nuclear +
bulge + disk) magnitude spans a much narrower range in absolute
magnitude and any correlation with radio emission is consequently much
weaker. 

For the Seyfert 2, the partial correlation coefficient is
0.13 which is significant at the 98.7\% level.  

The lack of a highly significant radio-optical luminosity correlation for
both classes of Seyferts is not altogether surprising, because the
radio emission is very likely linked to the presence of an active
nucleus (or starburst) while the optical emission has significant
contribution from stars in the bulge and disk of the underlying
galaxy. In fact, recent work indicates that the nuclear component may
contribute as little as 0.01\% of the {\it optical} continuum emission
of Seyfert galaxies (Ho \& Peng 2001). On the other hand, a
significant fraction of the radio emission from AGN likely originates
in the nuclear region.

\begin{figure}
\resizebox{\hsize}{!}{\includegraphics{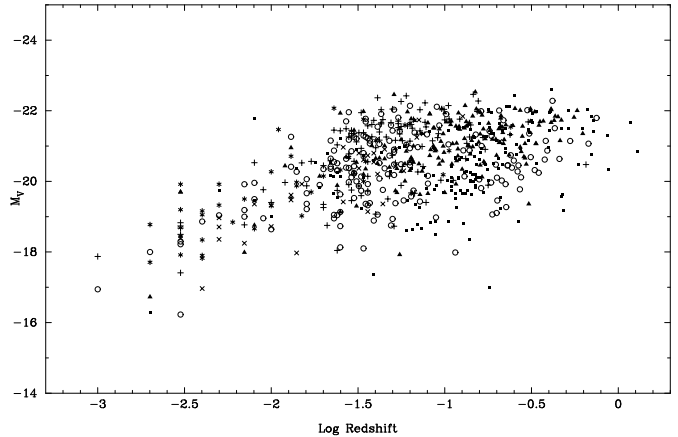}}
\caption{Absolute magnitude of the AGN in our sample as a function of redshift. The symbols are as in
\fig{lrzagn}.}
\labfig{abmzagn}
\end{figure}

\begin{figure}
\resizebox{\hsize}{!}{\includegraphics{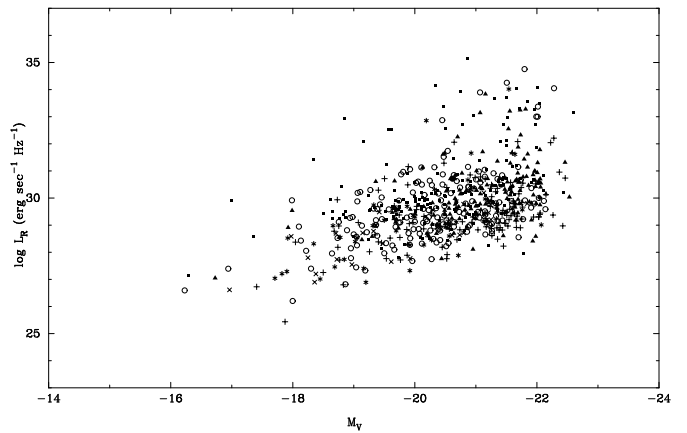}}
\caption{Radio
luminosity as a function of absolute magnitude. The symbols are as in \fig{lrzagn}.}
\labfig{lrabmagn}
\end{figure}

\section{Radio luminosity, source size and spectral index}

\labsecn{radlumseyfert} 

\subsection{Radio luminosity}

We have FIRST detections for 188 out of the 788 Seyfert 1 and 179 out
of the 298 Seyfert 2 galaxies in our sample. We have used the Seyfert
classifications listed in the VV01 catalog, without attempting to
verify them independently. The radio detection rate is thus about 25\%
for the Seyfert 1 and 60\% for the Seyfert 2 at the 1 $\mjy$
level. 

We list in \tab{meanseyfert} the mean radio luminosity for the two
classes of Seyferts. Errors listed are the root mean square deviations in the radio luminosity. The mean radio luminosity of both types of
Seyferts is two orders of magnitude lower than even the optically
selected quasars, most of which fall in the category of radio quiet
quasars. Though the radio emission from Seyferts is weak, it is still
two orders of magnitude higher than the typical emission from normal
galaxies (Wunderlich, Klein \& Wielebinski 1987). The table indicates
that the mean radio luminosity of the Seyfert 1 is greater than that
of the Seyfert 2, but this difference in mean radio luminosity is likely
caused by a bias in redshift -- the Seyfert 2 have a lower mean
redshift ($z_{mean}=0.11$) compared to the Seyfert 1 galaxies
($z_{mean}=0.18$). Such a distance-bias has been noted previously
(Dahari and de Robertis 1988; Roy et al. 1994). It arises in all
optically selected Seyfert samples since Seyfert 2's are less luminous
optically than Seyfert 1. However the standard deviation of the radio
luminosity is higher than the difference in mean luminosities and
therefore a strong statistical conclusion cannot be drawn from our
data. A Gehan-Wilconox test reveals that there is a 1\% probability that the
Seyfert 1 and Seyfert 2 are drawn from the same parent
population. This seems to indicate that the two Seyfert populations are drawn from different parent populations or at least have a strong selection bias.

It should be noted that a significant fraction of Seyfert galaxies in
the FIRST area only have an upper limit on their radio flux and radio
luminosity. When we wish to estimate the mean for a quantity that
includes both known values and upper limits, the Kaplan-Meier
estimator may be used. This estimator requires that the censoring be
random. Two factors contribute to making the censoring in radio luminosity random: 1. our
radio observations are of a sample of Seyfert galaxies selected mostly
using observations at non-radio wavelength and the radio and optical
luminosities are not corelated and 2. the galaxies we are considering
lie at different distances contributing to an increased randomness in
radio luminosity based censoring. 

We list in \tab{meanseyfert} the
mean-radio luminosity of (detections+non-detections) for the Seyfert 1
and Seyfert 2 galaxies. When non-detections are accounted for, the K-M
estimator predicts a slightly higher radio luminosity for the Seyfert
1 relative to the Seyfert 2. The error listed is that in the estimate of the mean, obtained using the Kaplan-Meier estimator. 

Ulvestad and Ho (2001) obtain similar results both with regard to
Seyfert 1's being more luminous than Seyfert 2's (although their
sample is not distance biased) and the results of the Gehan-Wilcoxon
test. They suggest that this difference in radio luminosities can be
accommodated in the unified scheme provided a minimum level of Seyfert
activity is required for the radio source to emerge from the vicinity
of the active nucleus. For our sample, the redshift bias (Seyfert 1
have higher mean redshift) is sufficient to explain the small
difference in mean radio luminosities.

\subsubsection{Normalized radio luminosities}

Edelson (1987) found no significant difference between the radio
luminosities of Seyfert 1 and Seyfert 2 galaxies in their
sample. Nevertheless, when radio luminosities were normalized by the
total optical luminosity, Seyfert 2 galaxies were found to be, on the
average, twice as luminous as Seyfert 1 galaxies, at the 99\%
confidence level. We used the radio to optical luminosity ratio $R$ as
a measure of normalized radio luminosities for the the two types of
Seyferts. For Seyfert 1 galaxies the mean value of $ \log R $ is \(
0.95\pm 1.00 \), while for Seyfert 2 galaxies, the mean value of \(
\log R \) is \( 0.73\pm 1.19 \). In fact, we find that the mean
normalized radio luminosity of the Seyfert 1 is somewhat larger than
that of the Seyfert 2. However, the small difference in mean normalized radio luminosity is almost certainly due to the optical obscuration of Seyfert 2 nuclei and is not intrinsic.

\subsection {Radio source sizes}

An expectation from the unified scheme is that radio source sizes of
Seyfert 1's should be systematically smaller than those of Seyfert 2's
due to foreshortening effects. Of course, the measured source size
will in general depend on the total radio flux of the source; this
effect can be roughly accounted for by using the FWHM of the radio source as a
measure of the size of the radio source. We used the deconvolved major
axis of the FIRST radio source as a measure of radio source size.

Another complication is that a significant number of our radio
detections of Seyferts (16\%) are unresolved in the FIRST survey and
are listed with a deconvolved major axis source size of 0.00 arcsec in
the FIRST catalog. 45 out of 188 Seyfert 1 and 22 out of 179 Seyfert 2
galaxies show unresolved radio emission. Additionally, the uncertainty
in radio source size depends upon the total flux of the source. Also,
the beam size varies as a function of the position of the source in
the sky, so the upper limits on the sizes of unresolved sources are
not uniform. In such a situation, it is nearly impossible to measure
radio source sizes reliably. Nevertheless, we use the K-M estimator to
compare the source sizes of Seyfert 1 and Seyfert 2 galaxies. We used
the FIRST beam size of 5.4 arcsec as the upper limit for sources that
were unresolved. The mean of the radio source
sizes was calculated using the Kaplan-Meier estimator, which gave
$11.931\pm1.949$ kpc and $5.794 +/- 0.572$ kpc for Seyfert 1's and
Seyfert 2's respectively. The error listed is the error in the estimate of the mean, obtained using the Kaplan-Meier estimator. Prima facie, this seem to run counter to the
predictions of the unified scheme. But the same subtle biases in
redshift and luminosity may also operate here. Perhaps more
importantly, the angular resolution of the FIRST data is insufficient
to measure sizes of small scale radio jets of the kind that operate in
radio-weak AGN, except in the nearest Seyferts. Also, the resolved emission in the FIRST survey in some cases, may be caused by galaxy-scale star formation, rather than AGN emission.

\subsection{Radio Spectral Index}

Radio flux measurements at 6 cm, as listed in VV01 were used to
compute radio spectral index for the Seyfert galaxies. Note that only
a few galaxies have 6 cm fluxes available. Also in general, the 20 cm
(FIRST) and 6 cm measurements are not made with ``matched beams''. This
means that data at the two wavelengths may well be probing emission over
different parts of the galaxy. Any conclusions that one may draw from
measurements of spectral indices are therefore subject to considerable
uncertainty.

The mean spectral index of Seyfert 1 is $-0.02\pm1.00$ (assuming $L_{\nu}
\propto \nu^{+\alpha}$) and that of Seyfert 2 is $0.15\pm1.44$.  The error estimates are the root mean square deviation of the spectral index. There
is a slight indication that the Seyfert 2 have larger radio spectral
indices than the Seyfert 1. But the difference between the spectral
indices of the two populations as measured by a Gehan-Wilcoxon gives
a 77\% probability that the two populations are the same.

\section{Conclusions}

The main results of this paper are: 

\begin{itemize}

\item We have detected nuclear radio emissions from 775 AGN of which
214 were previously undetected in the radio. $\sim27$\% of the known
AGN within the $\sim$9033 square degrees of sky covered by FIRST have
a flux higher than 1 $\mjy$.

\item We do not find a statistically significant correlation between
radio luminosity  of Seyfert 1 and Seyfert 2 and the optical luminosity of the
underlying host galaxy. The apparent correlation seen is induced
largely by the effects of redshift.

\item We find that mean radio luminosities of detected Seyfert 1 and
Seyfert 2 are consistent with expectations from the unified
scheme. When we compute the Kaplan Meier estimator for detections and non-detections for the two Seyfert types, they show a small difference in radio luminosity distribution. When normalised by optical luminosity, the radio luminosities of the two Seyfert populations are not significantly different. We
also find that radio source sizes in Seyfert 1 are significantly
larger than Seyfert 2, but this result is subject to numerous
caveats. The limited radio spectral index measurements indicate that
Seyfert 1 and Seyfert 2's are drawn from the same parent population.

\end{itemize}

\begin{acknowledgements}

The author wishes to thank Ajit Kembhavi for useful discussions and
the referee James Ulvestad for insighful comments that helped improve
this paper. The author's research at the IAP was supported
by the Indo-French centre for promotion of advanced scientific
research (CEFIPRA) through grants 1610-1 and 1910-1.
 
\end{acknowledgements}

\newpage

\begin{table}
\caption{Summary of radio detections}
\begin{tabular}{lc}
\hline \hline Number of VV01 AGN in FIRST area:&$\sim2840$\\
Number of AGN with radio detections:& 775\\
Number of non-detections:& $\sim 2065$\\
Percentage of AGN with radio emission &
\\
detected by FIRST:&27\%\\
\hline&
\\
\end{tabular}

\labtab{summagn}
\end{table}

\begin{table}
\caption[Number of radio detections of various classes of AGN]{Number of radio detections of various classes of AGN at the 1 \( \mjy  \)
flux limit of the FIRST survey. $^{a}$Many of these tentative identifications are from White
\etal (2000) and will require further spectroscopic observations to confirm their status. $^{b}$These sources 
do not have a class listed in the VV01 catalog.} 
\begin{tabular}{lll}
\hline\hline 
Class&Number in area&Number detected\\
&covered by FIRST&by FIRST\\
\hline&
&
\\
Seyfert 1&788&188\\
Intermediate type Seyferts&434&122\\
Seyfert 2&298&179\\
Seyfert 3 (LINERS)&63&41\\
Unconfirmed Seyferts$^{a}$&190&125\\
H-II regions&80&39\\
Unlisted$^{b}$ &989&81\\
\hline&&
\\
\end{tabular}

\labtab{summagn2}
\end{table}




\setcounter{table}{4}
\begin{table}
\caption{Radio luminosity, size and spectral index for Seyfert 1 and Seyfert 2 galaxies.}
\begin{tabular}{lcc}
\hline\hline&
Seyfert 1&
Seyfert 2\\
\hline&
&
\\
Mean log radio luminosity ($\ergsh$)&
30.24\( \pm 1.14 \) &
29.73\( \pm 1.41 \) \\
detections only&&\\
Mean log radio luminosity ($\ergsh$)&
28.71\( \pm 0.11 \) &
28.99\( \pm 0.11 \) \\
detections and upper limits&&\\
Mean radio source size (kpc)&
11.93\( \pm 1.95 \)&
5.79\( \pm 0.97 \)\\
Mean radio spectral index ($\alpha^{20}_6$)&
-0.02\( \pm 1.00 \)&
0.15\( \pm 1.44 \)\\
\hline&
&
\\
\end{tabular}
\labtab{meanseyfert}
\end{table}

\newpage
\begin{thebibliography}{}
\bibitem{} Becker, R. H., White, R. L., \& Helfand, D. J. 1995, ApJ,
 450, 559
 \bibitem{} de Bruyn, A. G \& Wilson, A. S. 1976, A\&A, 53, 93
 \bibitem{} de Bruyn, A. G \& Wilson, A. S. 1978, A\&A, 64, 433
 \bibitem{} Condon, J. J., Cotton, W. D., Greisen, E. W., Yin, Q. F., Perley,
 \bibitem{} R. A., Taylor, G. B. \& Broderick, J. J. 1998, AJ, 115, 1693
 \bibitem{} Dahari, O.~\& de Robertis, M.~M.\ 1988, ApJS, 67, 249
 \bibitem{} Edelson, R. A. 1987, ApJ, 313, 651
 \bibitem{} Ho, L.~C.~\& Peng, C.~Y. 2001, ApJ, 555, 650 
 \bibitem{} Roy, A.~L., Norris, R.~P., Kesteven, M.~J., Troup, E.~R., \& Reynolds, J.~E.\ 1994, ApJ, 432, 496
\bibitem{} Schmitt, H.~R., Antonucci, R.~R.~J., Ulvestad, J.~S., Kinney, A.~L., \& Pringle, J.~E. 2001, ApJ, 555, 663
\bibitem{} Ulvestad, J.~S.~\& Wilson, A.~S. 1984, ApJ, 285, 439
\bibitem{} Ulvestad, J.~S.~\& Wilson, A.~S. 1989, ApJ, 343, 659
\bibitem{} Ulvestad, J.~S.~\& Ho, L.~C. 2001, ApJ, 558, 561
\bibitem{} Veron-Cetty, M.P., \& Veron, P. 2001, A\&A, 374, 92
\bibitem{} Wadadekar, Y. \& Kembhavi, A. 1999, AJ, 118, 1435
\bibitem{} White, R. L., Becker, R. H., Helfand, D. J., \& Gregg, M. D. 1997, ApJ, 475, 479
\bibitem{} White, R.~L.~et al. 2000, ApJS, 126, 133
\bibitem{} Wunderlich, E., Wielebinski, R., \& Klein, U.\ 1987, A\&AS, 69, 487
\end{thebibliography}
\end{document}